\documentclass[aps,prb,reprint,superscriptaddress,showpacs]{revtex4-1}
\usepackage{graphicx}
\usepackage{mathrsfs} 
\usepackage[mathscr]{euscript}
\usepackage{amsmath,amssymb}
\usepackage{bm}
\usepackage{color}

\def\vec#1{ {\bm{#1}} }

\begin{document}


\title{Symmetry-protected topological order and negative-sign problem for SO($N$) bilinear-biquadratic chains}


\author{Kouichi Okunishi}
\affiliation{Department of Physics, Niigata University, Niigata 950-2181, Japan}
\author{Kenji Harada}
\affiliation{Graduate School of Informatics, Kyoto University, Kyoto 606-8501, Japan}
%


\date{\today}

\begin{abstract}
Using a generalized Jordan-Wigner transformation combined with the defining representation of the SO($N$) spin, we map the SO($N$) bilinear-biquadratic(BLBQ) spin chain into the $N$-color bosonic particle model.
We find that, when the Jordan-Wigner transformation disentangles the symmetry-protected topological entanglement, this bosonic model becomes negative-sign free in the context of quantum Monte-Carlo simulation.
For the SO(3) case, moreover, the Kennedy-Tasaki transformation for the $S=1$ BLBQ chain, which is also a topological disentangler, derives the same bosonic model through the dimer-R bases. 
We present the temperature dependence of energy, entropy and string order parameter for the SO($N=3, 4, 5$) BLBQ chains by a world-line Monte-Carlo simulation for the $N$-color bosonic particle model.
\end{abstract}

\pacs{75.10.Pq, 02.70.Ss, 75.10.Kt}

\maketitle
\section{Introduction}
The negative-sign problem in quantum Monte Carlo (QMC) simulations for frustrated spin systems or fermion systems  has been a longstanding issue in quantum many-body physics.
Transition probabilities of world-line configurations in QMC simulations are usually constructed via product states relying on single particle bases, where the nontrivial entanglement of the world lines often generates negative weights. 
In general, it is known that the negative-sign problem is NP hard\cite{Troyer}, which implies that a mere use of a basis change of a local Hilbert space is not able to settle the problem. 
In other words, it is suggested that the nonlocal quantum entanglement structure due to the frustration effect may play an essential role in the negative-sign problem.

Of course, a general solution for the negative-sign problem is very tough.
 As for one-dimensional(1D) systems, however, two nontrivial examples where the intrinsic negative sign has been removed are known: the $S=1/2$ zigzag spin ladder\cite{Nakamura} and the SU($N$) spin chain\cite{Mila,Frischmuth}. 
For the former case,  the negative sign was removed by the so called dimer-R basis\cite{Nakamura}  combined with a nonlocal unitary transformation similar to the Kennedy-Tasaki(KT) transformation for the $S=1$ valence-bond-solid(VBS) state\cite{KT,Kennedy,Kohmoto-Tasaki}.
A key point is that  the KT-type transformation enables  us to rewrite the Hamiltonian into a kind of ferromagnetic system.
In addition, it should be remarked that the dimer-R basis in Ref. [\onlinecite{Nakamura}] can be viewed as the maximal-entangled pair in quantum information terminology.
In the modern view point, the KT transformation disentangles the topological entanglement of the dimer groundstate of the zigzag ladder, suggesting that the entanglement of the groundstate is certainly related to the negative-sign problem in a class of quantum spin systems.
For the latter example, i.e. the SU($N$) spin chain, the negative sign can be removed by a generalized Jordan-Wigner transformation.\cite{Mila} 
However, the situation is slightly different from the zigzag ladder.
It can be exactly solved by the Bethe ansatz, where the groundstate is gapless\cite{Sutherland}, but  the role of the Jordan-Wigner transformation has been unclear from the entanglement viewpoint.
In order to systematically control the negative sign in 1D quantum systems, it is necessary to construct a unified theory for the above nonlocal transformations, which may provide some important hints for addressing the problem in the 1D frustrated quantum spin systems.

For the purpose of revealing the relation between negative sign and entanglement, the most suitable target is the $S=1$ bilinear biquadratic (BLBQ) chain, which includes both of the Affleck-Kennedy-Lieb-Tasaki(AKLT) model, the groundstate of which is exactly described by the VBS state,\cite{AKLT} and the SU(3) chain that is solved by the Bethe ansatz\cite{Sutherland}.
In the context of the QMC, the negative sign appears for the BLBQ chain in the parameter region including the AKLT and SU(3) points.
Thus, we can  investigate the connection of the KT transformation and the generalized Jordan-Wigner transformation on an equal footing.
Moreover, the hidden $Z_2\times Z_2$ symmetry of the Haldane phase and the associated entanglement spectrum  recently attract renewed interest as a striking example of the symmetry-protected topological order\cite{Wen,Pollmann}, although it has been a long time since the Haldane conjecture.\cite{Haldane}
Thus, it is also worthwhile to understand how the hidden symmetry can be related to the sign problem.

In this paper, we first discuss the relation between the negative-sign problem and the nonlocal transformations for the $S=1$ BLBQ chain in detail, and then generalize it to the hidden topological order of the SO($N$) BLBQ chains\cite{Xiang1,Xiang2}.
In Sec. 2, we particularly find that the two distinct approaches of the KT transformation combined with the dimer-R basis and a generalized Jordan-Wigner transformation on the defining representation of the SO(3) group generate the same negative-sign-free Hamiltonian in the parameter region of the Haldane phase.
Moreover, the negative-sign-free Hamiltonian can be represented as the scattering diagrams of bosonic particles carrying three types of colors, which are suitable for a world-line QMC simulation based on the directed-loop algorithm\cite{Sandvik,Harada::2004}.
In Sec. 3, we straightforwardly generalize the theory to the SO($N$) BLBQ model, where the negative sign can also be removed in the parameter region corresponding to the VBS-type groundstate by the generalized Jordan-Wigner transformation as well. 
Also, the  SO($N$) version of the VBS state can be disentangled to the classical product state by this transformation. 
In Sec. 4, we next demonstrate QMC simulations for negative-sign-free Hamiltonians generated from  SO(3), SO(4), and SO(5) chains.
The temperature dependence of various observables and  string-correlation functions is presented. 
In Sec. 5, we summarize results and discuss further perspectives.

\section{$S=1$ bilinear-biquadratic chain}

\subsection{Hamiltonian}

Let us start with the $S=1$ BLBQ chain, which has been extensively studied in the connection with the Haldane state of the $S=1$ Heisenberg chain\cite{Solyom,Nomura,Fath,Xiang0,Schollweck,Itoi}.
Here, we focus on the essential physics of the hidden symmetry and entanglement associated with the negative-sign problem.
The Hamiltonian of the $S=1$ BLBQ chain is given by
\begin{equation}
{\cal H} =\sum_i h_{i,i+1}, 
\end{equation}
with the local interaction term defined as 
\begin{equation}
{h}_{i,i+1}=  \vec{S}_i\cdot \vec{S}_{i+1}+ \alpha [(\vec{S}_i\cdot \vec{S}_{i+1})^2-1  ] ,
\label{S1H}
\end{equation}
where $\vec{S}$ is the standard $S=1$ spin matrix in the $S^z$-diagonal basis.
We basically consider the open boundary condition.
This Hamiltonian includes a series of important models: 
$\alpha=0$ is the $S=1$ Heisenberg antiferromagnetic chain and $\alpha=1/3$ is the AKLT model, the groundstate of which is exactly described by the VBS state\cite{AKLT}. 
Recently, the VBS/Haldane state is often refered to as a typical example of the  symmetry-protected topological order.\cite{Wen,Pollmann}
Moreover,  $\alpha=1$ corresponds to the  SU(3) chain in the fundamental representation, which can be solved exactly by the Bethe ansatz\cite{Sutherland}, and $\alpha=\infty$ corresponds to the SU(3)-singlet chain in the [1,1] representation, which has Temperly-Lieb equivalence to the 9-states quantum Potts model\cite{Batchelor}.
Also, $\alpha=-1$ is an integrable point, which is Bethe-ansatz solvable,  with a gapless groundstate\cite{TB}. 
Of course, the properties of these models have been clarified by a variety of analytic and numerical approaches, and thus they are very useful to see the connection between the hidden symmetry and the negative-sign problem.

The explicit matrix elements of ${h}_{i,i+1}$ are written as
\begin{equation}
h_{i,i+1}=
\left( \begin {array}{ccccccccc} 
1&0&0&0&0&0&0&0&0\\ 
0&0&0&1&0&0&0&0&0\\ 
0&0&-1+\alpha&0&1-\alpha&0&\alpha&0&0\\
0&1&0&0&0&0&0&0&0\\ 
0&0&1-\alpha&0&\alpha&0&1-\alpha&0&0\\ 
0&0&0&0&0&0&0&1&0\\ 
0&0&\alpha&0&1-\alpha&0&-1+\alpha&0&0\\ 
0&0&0&0&0&1&0&0&0\\ 
0&0&0&0&0&0&0&0&1
\end {array} \right)
\end{equation}
which contains positive off-diagonal elements resulting in the negative-sign problem for $\alpha>0$
(On the other hand, there is no negative-sign problem in $\alpha \le 0$\cite{Harada::2001, Harada::2002}).
This negative sign cannot be removed by such a local unitary transformation as  $\pi$-rotation of spin at every second site, so that  QMC simulations have  not been effective for investigating the VBS groundstate so far.

\subsection{nonlocal transformations and diagrammatic representation}

In order to remove the negative sign, the KT transformation plays a crucial role. 
The  KT transformation maps the non-local string order\cite{Nijs} to a kind of the ferromagnetic order, which can be viewed as the classical state where the all spins are disentangled with each other\cite{KT,Kennedy,Oshikawa,Okunishi}.
Thus, it may be expected that the KT transformation reduces the negative-sign problem as in the case of the usual ferromagnetic chain where no negative sign appears.
The KT transformation was originally defined as sequential spin and sign flips based respectively on the hidden antiferomagnetic order, and on the number of ``0" spins existing to the left of a certain site.
However, it is more useful to rewrite the KT transformation as a product over the pair disentanglers,\cite{Okunishi,Oshikawa}
\begin{equation}
{\cal U} = \prod_{\langle i, j \rangle} U_{i,j} 
\end{equation}
where the pair disentangler is defined for $i<j$ as 
\begin{equation}
U_{i,j}=e^{i\pi S_i^zS_j^x},
\end{equation}
and $\langle i,j\rangle$ runs over all spin pairs in the chain.
Applying the KT transformation to the BLBQ chain,  we have 
\begin{eqnarray}
&& {\cal U} h_{i,i+1} {\cal U }^\dagger =\nonumber \\
&&\left( \begin {array}{ccccccccc} 
-1+\alpha&0&0&0&-1+\alpha&0&0&0&\alpha\\ 
0&0&0&-1&0&0&0&0&0\\ 
0&0&1&0&0&0&0&0&0\\
0&-1&0&0&0&0&0&0&0\\ 
-1+\alpha&0&0&0&\alpha&0&0&0&-1+\alpha\\ 
0&0&0&0&0&0&0&-1&0\\ 
0&0&0&0&0&0&1&0&0\\ 
0&0&0&0&0&-1&0&0&0\\ 
\alpha&0&0&0&-1+\alpha&0&0&0&-1+\alpha
\end {array} \right) .
\label{KTH}
\end{eqnarray}
In this equation, most of the positive off-diagonal elements were actually removed, but a source of the negative sign still remains at (1,9) and (9,1).

Inspired by the dimer-R basis for the zigzag ladder\cite{Nakamura}, we further introduce a local basis change as follows,
\begin{eqnarray}
\left( \begin {array}{c} 
|1\rangle\\
|2\rangle\\
|3\rangle
\end {array} \right)
=
\left( \begin {array}{ccc} 
\frac{1}{\sqrt{2}} & 0 & -\frac{1}{\sqrt{2}}\\
\frac{1}{\sqrt{2}} & 0 & \frac{1}{\sqrt{2}}\\
0 & 1 & 0 
\end {array} \right) 
\left( \begin {array}{c} 
|+\rangle\\
|0\rangle\\
|-\rangle
\end {array} \right) 
\label{dimerR}
\end{eqnarray}
which is equivalent to the transformation found in Ref.[\onlinecite{Kennedy}].
In the following, we call the label of the ket state $|n\rangle$ the ``color''.
Writing this local transformation matrix of Eq. (\ref{dimerR}) for $i$th site as $V_i$, we define ${\cal V}=\prod_i V_i$ for the entire chain.
We then apply ${\cal V}$ to the Hamiltonian (\ref{KTH}) to obtain
\begin{equation}
\tilde{\cal H} = {\cal VU} {\cal H}{\cal U}^\dagger{\cal V}^\dagger,
\end{equation}
in which the explicit form of the local Hamiltonian is
\begin{eqnarray}
&&\tilde{h}_{i,i+1}\equiv\nonumber \\
&&\left( \begin {array}{ccccccccc}
 \alpha&0&0&0&-1+\alpha&0&0&0&-1+\alpha\\ 
0&0&0&-1&0&0&0&0&0\\ 
0&0&0&0&0&0&-1&0&0\\
0&-1&0&0&0&0&0&0&0\\ 
-1+\alpha&0&0&0&\alpha&0&0&0&-1+\alpha\\ 
0&0&0&0&0&0&0&-1&0\\ 
0&0&-1&0&0&0&0&0&0\\ 
0&0&0&0&0&-1&0&0&0\\ 
-1+\alpha&0&0&0&-1+\alpha&0&0&0&\alpha
\end {array} \right) .
\label{S1DR}
\end{eqnarray}
In this Hamiltonian,   all the  off-diagonal elements become non-positive for $\alpha\le 1$, so that the system can be reduced to a ferromagnetic chain, where the negative-sign problem does not appear.
From the entanglement viewpoint, all sites in the groundstate are disentangled from each other\cite{KT,Kennedy, Okunishi} and thus there is no topological entanglement.
This fact should be contrasted with the Haldane state for the original Hamiltonian which carries the topological entanglement entropy associated with the protected $Z_2$ symmetry, $S_{\rm EE}=\ln 2$, for the open chain.
In this sense, a class of negative-sign problem in one dimension may share the same background with the symmetry-protected topological entanglement.

\begin{figure}[htb]
\centering
\resizebox{6cm}{!}{\includegraphics{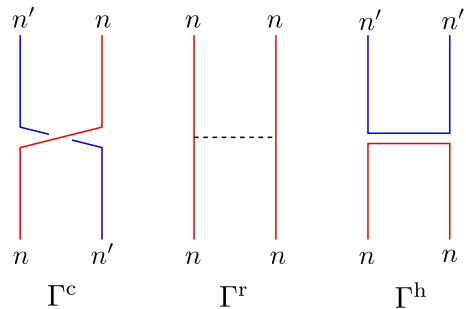}}
\caption{Diagrammatic representation of the interaction terms in the Hamiltonian (\ref{tildegamma}). The indices $n$ and $n'$ represent colors of particles.
The superscripts of $\Gamma$ indicate abbreviations of cross, repulsive and horizontal, respectively}

\label{diagram}
\end{figure}

An essential point in Eq. (\ref{S1DR}) is that the roles of the matrix elements  can be classified by three types of the interaction terms,
\begin{equation}
\tilde{h}_{i,i+1} = - \Gamma^{\rm c}_{i,i+1} + \alpha \Gamma^{\rm r}_{i,i+1} - (1-\alpha)\Gamma_{i,i+1}^{\rm h}.
\label{tildegamma}
\end{equation}
In this equation, $\Gamma^{\rm c}$ represents the exchange of bosonic particles carrying different colors between the $i$th and the $i+1$th sites, and $\Gamma^r$ represents the repulsive interaction ($\alpha>0$) between particles of the same color, corresponding to the diagonal elements in  Eq. (\ref{S1DR}).
Finally, the pair creation and annihilation of different color particles are denoted as $\Gamma^h$. (The  matrix elements of $\Gamma$s  will be explicitly given by Eqs. (\ref{su3algebra1}), (\ref{su3algebra2}) and (\ref{su3algebra3}) ).
In Fig. \ref{diagram}, moreover, we illustrate these terms as schematic diagrams of the world lines,  which play an important role in the world-line QMC simulation.

Here it should be recalled that,  at $\alpha=1$,  the symmetry is enhanced to SU(3), where the pair annihilation and creation term vanishes in Eq. (\ref{tildegamma}).\cite{TLalgebra}
Taking account of this property, we can further rewrite the Hamiltonian $\tilde{h}_{i,i+1}$ as 
\begin{equation}
\tilde{h}_{i,i+1}= \tilde{h}^{[1,0]}_{i,i+1} -(1-\alpha){h}_{i,i+1}^{[1,1]}
\end{equation} 
with
\begin{eqnarray}
\tilde{h}_{i,i+1}^{[1,0]}& =&  -\Gamma_{i,i+1}^{\rm c} + \Gamma_{i,i+1}^{\rm r},\\
{h}_{i,i+1}^{[1,1]} &=&  \Gamma_{i,i+1}^{\rm h} + \Gamma_{i,i+1}^{\rm r}.
\end{eqnarray}
where $\tilde{h}_{i,i+1}^{[1,0]}$ can be associated with the SU(3) spin Hamiltonian in the [1,0] fundamental representation without the sign in front of $\Gamma^{\rm c}$ term, and ${h}_{i,i+1}^{[1,1]}$ describes the SU(3)-singlet interaction in the [1,1] representation. 
Thus, it can be expected that the $\Gamma$ terms are also closely related to the generators of the SU(3) algebra.
In fact, it is found that $\Gamma$s are directly written as 
\begin{eqnarray}
\Gamma_{i,i+1}^{\rm c}&=&\sum_{\mu\ne\nu} S^{\mu\nu}_iS^{\nu\mu}_{i+1},\label{su3algebra1}\\
\Gamma_{i,i+1}^h& =& \sum_{\mu\ne \nu} S^{\mu\nu}_iS^{\mu\nu}_{i+1}, \label{su3algebra2}
\\
\Gamma_{i,i+1}^r&=& \sum_\mu S_i^{\mu\mu}S_{i+1}^{\mu\mu},\label{su3algebra3}
\end{eqnarray}
where $S^{\mu\nu}$($\mu,\nu = 1, 2, 3$) are the $3\times 3$ matrices generating the SU(3) algebra.
Explicitly, $S^{\mu\nu}$ obey the SU(3) commutation relation $[S^{\mu\nu}, S^{\mu'\nu'}]=\delta_{\nu,\mu'}  S^{\mu\nu'} - \delta_{\mu,\nu'} S^{\mu'\nu}$.
Moreover, if we introduce the Schwinger boson $b_\mu$ and write the matrices as  $S^{\mu\nu}=b_\mu^\dagger b_\nu $  with the constraint $\sum_\mu b_\mu^\dagger b_\mu =3$,  we can view the diagrams in Fig. \ref{diagram} as the scattering processes of the 3-color bosonic particles.

\subsection{generalized Jordan-Wigner transformation and the defining representation}

As for the propose of settling the negative-sign problem of the $S=1$ BLBQ chain, Eqs (\ref{S1DR}) or (\ref{tildegamma}) may be sufficient.
For the generalization to the SO($N$) case, however, it is essential to reveal the relation of $\tilde{h}^{[1,0]}$ to the standard SU(3) spin Hamiltonian.
Let us recall that the negative sign of the  SU($N$) spin chain is removed by a generalized Jordan-Wigner transformation\cite{Mila}.
Here, we construct the transformation matrix as a product form of the following pair operator, 
\begin{equation}
Q_{i,j} \equiv  {\rm diag}(1,-1,-1,1,1,-1,1,1,1),
\end{equation}
where ${\rm diag}(\cdots)$ represents a diagonal matrix whose diagonal elements are ``$\cdots$''.
Similarly to the KT transformation,  we can write the generalized Jordan-Wigner transformation for the entire chain as 
\begin{equation}
{\cal Q} = \prod_{\langle i,j \rangle} Q_{i,j}
\label{JW}
\end{equation}
where the product is taken for the all pairs of spins in the open chain\cite{Okunishi}.   
This operator ${\cal Q}$ inverts the sign of a matrix element, if particles of different colors are exchanged.
Although the Jordan-Wigner transformation is usually defined as sequential sign flips, the product form of Eq. (\ref{JW}) is more useful in practical sense.
We list some important properties of ${\cal Q}$ below.
\begin{itemize}
\item  $[Q_{i,j},Q_{k,l}]=0$ for any $i,j,k,l$. 
Thus, the order of operators in  ${\cal Q}$ is irrelevant.

\item ${\cal Q}^2=1$. Thus, ${\cal Q}^{-1}={\cal Q}$.

\item For any adjacent sites,   ${\cal Q}\Gamma^{\rm c} {\cal Q}= -\Gamma^{\rm c}$, ${\cal Q} \Gamma^{\rm r} {\cal Q}= \Gamma^{\rm r}$, and ${\cal Q} \Gamma^{\rm h} {\cal Q}= \Gamma^{\rm h}$.
\end{itemize}
Since $Q_{i,j}$ is a diagonal matrix, the proofs are  straightforward.

Using the properties above, we can easily show that  
\begin{equation}
{h}_{i,i+1}^{[1,0]}\equiv {\cal Q}\tilde{h}_{i,i+1}^{[1,0]}{\cal Q} = \Gamma_{i,i+1}^{\rm c} + \Gamma_{i,i+1}^{\rm r} ,
\end{equation}
which is just the nearest-neighbor interaction of the  SU(3) chain.
This is basically the same as the negative sign vanishing mechanism for the SU($N$) chain in Ref.[\onlinecite{Mila}].
On the other hand, we find that the SU(3)-singlet part is invariant under the Jordan-Wigner transformation: ${h}^{[1,1]}_{i,i+1}= {\cal Q}{h}^{[1,1]}_{i,i+1}{\cal Q}$.
We therefore arrive at 
\begin{equation}
\hat{\cal H} = \sum_i \hat{h}_{i,i+1} \equiv {\cal Q} \tilde{\cal H} {\cal Q} 
\end{equation}
where
\begin{eqnarray}
\hat{h}_{i,i+1}&\equiv&  {\cal Q} \tilde{h}_{i,i+1} {\cal Q}\nonumber \\
&=&{h}_{i,i+1}^{[1,0]} -(1-\alpha){h}_{i,i+1}^{[1,1]} \nonumber \\
            &=& \Gamma_{i,i+1}^{\rm c} + \alpha \Gamma_{i,i+1}^{\rm r} - (1-\alpha)\Gamma_{i,i+1}^{\rm h}.
\label{hath}
\end{eqnarray}

At the present stage, $\hat{\cal H}$ is a descendant of the sequence of the transformations, ${\cal U}$, ${\cal V}$ and  ${\cal Q}$.
In particular,  ${\cal U}$ and  ${\cal Q}$ involve the similar nonlocality.
In addition, the Hamiltonian $\hat{h}_{i,i+1}$ maintains very simple structure, although the negative sign recovered by ${\cal Q}$.
Thus, one may expect a more direct relation between the original Hamiltonian ${\cal H}$ and the descendant Hamiltonian $\hat{\cal H}$, which is actually the case. 
In order to clarify the relation, we introduce another local unitary transformation matrix,
\begin{equation}
R = \left( \begin {array}{ccc} 
-i/\sqrt {2}&0&i/\sqrt {2}\\ 
1/\sqrt {2}&0&1/\sqrt {2}\\ 
0&i&0\end {array} \right),
\end{equation}
and write the transformation for the entire chain as ${\cal R}\equiv \prod_i R_i$.
Then, it is straightforward to see
\begin{equation}
\hat{\cal H}={\cal R} {\cal H} {\cal R}^\dagger .
\end{equation}
Moreover, we find that the $S=1$ spin matrices are transformed by $R$ as follows
\begin{eqnarray}
L^x &\equiv& R S^x R^\dagger = -i(S^{23}-S^{32}),\nonumber \\
L^y &\equiv& R S^y R^\dagger = -i(S^{31}-S^{13}),\nonumber \\
L^z &\equiv& R S^z R^\dagger= -i(S^{12}-S^{21}),
\label{spindefrep}
\end{eqnarray}
where $S^{\mu\nu}$ are the 3$\times 3$ matrices of the SU(3) algebra.
An important point is that the new spin matrices $\vec{L}$ are nothing but the generators of the SO(3) rotational group in the defining representation.
In other words,  $\hat{h}_{i,i+1}$ is written as   
\begin{equation}
\hat{h}_{i,i+1}= \vec{L}_{i}\cdot \vec{L}_{i+1} +\alpha [( \vec{L}_{i}\cdot \vec{L}_{i+1})^2-1].
\end{equation}
which is just the BLBQ interaction in the defining representation of the SO(3) rotational group.

\begin{figure}[htb]
\centering
\resizebox{8.5cm}{!}{\includegraphics{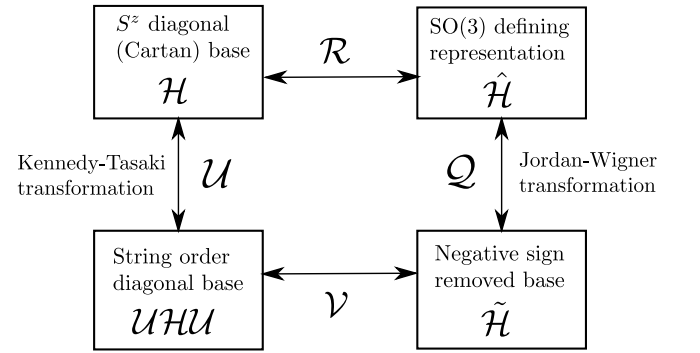}}
\caption{The relations among the various representations and transformations. 
The transformations in the vertical direction are nonlocal, while those in the horizontal direction are local.}
\label{relation}
\end{figure}

In the defining representation,  the negative sign of the BLBQ chain can be directly removed by the generalized Jordan-Wigner transformation (\ref{JW}), instead by the KT transformation for the chain in the $S^z$-diagonal base.
Now, we can summarize the relations closed among the various representations and transformations in Fig. \ref{relation}.

Here, we should comment on the relation between the two nonlocal transformations, ${\cal U}$ and ${\cal Q}$.
In order to see it, we rewrite the KT transformation in the defining representation,
\begin{eqnarray}
{\cal R}U_{i,j}{\cal R}^{\dagger}&=&e^{i\pi L^{z}_iL^{x}_j} \nonumber \\
     &=& {\rm diag}(1,-1,-1,1,-1,-1,1,1,1)
\end{eqnarray}
for $\forall i,j$, which is simplified to a diagonal matrix.
Then, we can see that the difference between ${\cal U}$ and the Jordan-Wigner transformation ${\cal Q}$ is that ${\cal U}$ has a phase factor $(-1)^{q_{22}}$ compared to ${\cal Q}$, where $q_{22}$ is the  number of all ``2-2" pairs in the entire chain. 
Thus, the difference between ${\cal U}$ and ${\cal Q}$ is of course nonlocal.
However, these two nonlocal transformations can be bridged by the local unitary transformations ${\cal V}$ and ${\cal R}$,
\begin{eqnarray}
{\cal R}U_{i,j}{\cal R}^{\dagger} = {\cal V}^\dagger {\cal Q}_{i,j} {\cal V},
\end{eqnarray}
which suggests the nontrivial symmetry involved in the BLBQ chain.

As is seen in Eq. (\ref{spindefrep}), the three rotation axis can be equivalently treated in the defining representation,  in contrast to the standard $S^z$-diagonal representation where the $z$ axis plays the special role of the quantization axis.
Indeed, the order of the coloring label, $n$  in Eq. (\ref{dimerR}) is irrelevant to the subsequent results.
This implies that the treatment of the dihedral group symmetry $D_2$ becomes manifest in the defining representation\cite{Pollmann}. 
Thus, one can expect that the negative-sign-free basis can be more systematically generalized, on the basis of the defining representation of the SO($N$) group.

\subsection{correlation functions}

We next examine how the correlation functions are transformed through the two paths in Fig. 2.
For later convenience, we define real symmetric matrices $\vec{T}$,
\begin{eqnarray}
T^x \equiv S^{23}+S^{32},\\
T^y \equiv S^{13}+S^{31},\\
T^z \equiv S^{12}+S^{21},
\end{eqnarray}
which can be considered as some kind of dual matrices to the SO(3) spins  $\vec{L}$.
For the path of the Jordan-Wigner transformation ${\cal Q}$,  we have
\begin{eqnarray}
{\cal Q} L_i^x {\cal Q}&=& e^{i\pi \sum_{j<i}({L}_j^y+1)} {L}_i^x e^{i\pi \sum_{j>i}({L}_j^z+1)} \label{jwx}\\
{\cal Q} L_i^y {\cal Q}&=& e^{i\pi \sum_{j<i}{L}_j^z} {L}_i^y e^{i\pi \sum_{j>i}{L}_j^x}
\label{jwy}\\
{\cal Q} L_i^z {\cal Q}&=& e^{i\pi \sum_{j<i}({L}_j^x+1)} {L}_i^z e^{i\pi \sum_{j>i}({L}_j^y+1)}. \label{jwz}
\end{eqnarray}
While for the rout of the KT transformation combined with the dimer-R basis, the spin matrices are given by
\begin{eqnarray}
{\cal V}{\cal U} S_i^x {\cal U}{\cal V}^\dagger&=& {T}_i^x e^{i\pi \sum_{j>i}{T}_j^x} \label{ktrx}\\
{\cal V}{\cal U} S_i^y {\cal U}{\cal V}^\dagger&=& e^{i\pi \sum_{j<i}{T}_j^z} {L}_i^y e^{i\pi \sum_{j>i}{T}_j^x} \label{ktry}\\
{\cal V}{\cal U} S_i^z {\cal U}{\cal V}^\dagger&=&  e^{i\pi \sum_{j<i}{T}_j^z} {T}_i^z
\label{ktrz}
\end{eqnarray}
where we have used 
\begin{equation}
V_i\vec{S}_iV_i^\dagger = \vec{T}_i.
\end{equation}
Here, it should be noted that Eqs. (\ref{jwx}), (\ref{jwy}) and (\ref{jwz}) basically have a symmetric form with respect to $x$, $y$, $z$, in contrast to (\ref{ktrx}), (\ref{ktry}) and  (\ref{ktrz}).

Using the above results, we have the same expressions for the  correlation functions for both paths, 
\begin{eqnarray}
\langle S_i^a S_j^a\rangle_{\cal H} &=&\langle L_i^a L_j^a\rangle_{\hat{\cal H}}\nonumber\\
&=& -\langle  {T}_i^a e^{i\pi \sum_{i<k<j}{L}_k^a}{T}_j^a \rangle_{\tilde{\cal H} } \nonumber \\
&=& -\langle  {T}_i^a e^{i\pi \sum_{i<k<j}{T}_k^a}{T}_j^a \rangle_{\tilde{\cal H} }
\end{eqnarray}
where $a \in x, y, z$. 
Note that, for example,   $\langle \cdots \rangle_{\tilde{\cal H}}$ indicates the expectation value with respect to the basis of the Hamiltonian $\tilde{\cal H}$.
In the last equation, we have used the identity,
\begin{equation}
 e^{i \pi L^a}=e^{i \pi T^a},
\label{LTidentity}
\end{equation}
which is crucial to prove the equivalence between the Jordan-Wigner and KT transformations at the correlation function level.
We also illustrate that the string correlation functions are mapped as
\begin{eqnarray}
&&\langle S_i^ae^{i\pi\sum_{i<k<j}S^a_k} S_j^a\rangle_{\cal H} =\langle L_i^ae^{i\pi \sum_{i<k<j} L_k^a} L_j^a\rangle_{\hat{\cal H}}\nonumber \\
&& \qquad = -\langle  {T}_i^a{T}_j^a \rangle_{\tilde{\cal H} },
\label{string}
\end{eqnarray}
which is consistent with the known result of the KT transformation.\cite{Kennedy}

\subsection{AKLT point}

At the AKLT point, $\alpha=1/3$, the two site Hamiltonian (\ref{S1H}) becomes the projection operator to  $S_{\rm tot}=2$ sector of the composite spin, where the groundstate is described by the $2\times 2$ matrix product state (MPS) carrying the bi-partition entanglement entropy, $S_{\rm EE}=\ln 2$.\cite{MPSAKLT}
In the negative-sign-removed representation, however, the groundstate is a kind of classical ferromagnetic state, where every site is disentangled from the others\cite{KT,Kennedy, Okunishi} and thus there is no topological entanglement.
Then, the four degenerate groundstates of the AKLT point are written as the very simple form
\begin{equation}
|\Phi^\nu\rangle = \prod_i |\phi^\nu_i \rangle 
\label{classical_aklt}
\end{equation}
with 
\begin{eqnarray}
|\phi^1\rangle = \frac{1}{\sqrt{3}}(|1\rangle + |2\rangle + |3\rangle),\\
|\phi^2\rangle = \frac{1}{\sqrt{3}}(|1\rangle - |2\rangle + |3\rangle),\\
|\phi^3\rangle = \frac{1}{\sqrt{3}}(|1\rangle + |2\rangle - |3\rangle),\\
|\phi^4\rangle = \frac{1}{\sqrt{3}}(|1\rangle - |2\rangle - |3\rangle).
\end{eqnarray}
An important point is that the degenerating states are distinguished by the distribution of ``$-$'' in front of the three color kets, except for the overall sign.
Thus, the total number of the degeneracy is illustrated as the number of possible distributions of ``$-$'', i.e. $2^3  /2 = 4$, which is consistent with the $Z_2 \times Z_2$ symmetry breaking.
This fact is also  significant for the generalization to the SO($N$) spin chain.

\section{generalization to  SO($N$) chains}

We generalize the theory for the $S=1$ BLBQ chain to the SO($N$) BLBQ chains.
Let us start with the defining representation of the SO($N$) rotational group, which is given by the $N\times N$ matrix,\cite{Georgi}
\begin{equation}
(L^{ab})_{x,y}= -i (\delta_{a,x}\delta_{b,y} -\delta_{b,x}\delta_{a,y})
\label{sondef}
\end{equation}
where $L^{ab}$ is the generator of a rotation in the $ab$ plane, and $a, b=1\cdots N$. 
Also, the subscript $x,y(=1\cdots N)$ is the matrix index  representing color of a particle.
Note that $L^{ab}$ is antisymmetric with respect to the permutation of $a\leftrightarrow b$.
For the SO(3) case, we have explicitly $L^x=L^{23}$, $L^y=-L^{13}$ and  $L^z=L^{12}$.
If we take $L^z$ as a Cartan generator and use the $R$ matrix, we have the usual $S^z$-diagonal representation of the $S=1$ spin matrix.
However, we  directly deal with Eq. (\ref{sondef}) and then introduce the generalized Jordan-Wigner transformation rather than the KT one based on the Cartan-generator-diagonalizing bases.

The SO($N$) BLBQ Hamiltonian is written as\cite{Xiang1,Xiang2}
\begin{equation}
\hat{\cal H}= \sum_i \hat{h}_{i,i+1}
\end{equation}
with
\begin{equation}
\hat{h}_{i,i+1}= \sum_{b>a} L^{ab}_i L^{ab}_{i+1} + \frac{\alpha}{N-2}\left[  \left(\sum_{b>a} L^{ab}_i L^{ab}_{i+1}\right)^2 -1 \right]
\label{eq:SONBLBQ}
\end{equation}
where the parametarization $\frac{\alpha}{N-2}$ is later convenience.
Note that $\alpha=1$ is the SU($N$) point, and $\alpha=\frac{N-2}{N}$ is the SO($N$)-VBS point, where the groundstate is exactly written in the matrix product form.
In particular, the symmetry-protected topological order appears in a certain region of $\alpha <1$ (The lower bound of $\alpha$ depends on $N$).

As in the case of SO(3),   the matrix elements of the local Hamiltonian in the defining representation can be represented  as  
\begin{equation}
\hat{h}_{i,i+1}= \Gamma_{i,i+1}^{\rm c} + \alpha \Gamma_{i,i+1}^{\rm r} - (1-\alpha)\Gamma_{i,i+1}^{\rm h}
\label{sonh}
\end{equation}
where the $\Gamma$s are $N^2 \times N^2$ matrices  representing the scattering of the particles of $n_i$ and $ n_{i+1}$$(= 1,\cdots N)$ colors. 
 $\Gamma^{\rm c}$ denotes the particle exchange of different colors, $\Gamma^r$ indicates the repulsion of the same color particles, and $\Gamma^h$ means the pair creation and annihilation of different colors. 
The corresponding world-line diagrams are the same as in Fig. \ref{diagram}.
Their matrix elements can be explicitly given by the SU($N$) version of Eqs. (\ref{su3algebra1}), (\ref{su3algebra2}), and (\ref{su3algebra3}), with  the SU(3) generators replaced by those of SU($N$).
In the Hamiltonian (\ref{sonh}) of $\alpha\le 1$,  the negative sign  comes from $ \Gamma^{\rm c}$ term.

In order to invert the sign of  $ \Gamma^{\rm c}$, we define a generalized Jordan-Wigner transformation as
\begin{equation}
{\cal Q} = \prod_{\langle i,j \rangle} Q_{i,j}
\end{equation}
where the product is taken for all the pairs in the chain.
The  $N^2\times N^2$ diagonal matrix $Q_{i,j}$ is explicitly constructed for $i<j$ as 
\begin{eqnarray}
&& Q_{i,j} \equiv \nonumber\\
&& {\rm diag}(\overbrace{1}^1,\overbrace{-1,\cdots -1}^{N-1},\cdots, \overbrace{1,\cdots1}^{l},\overbrace{-1\cdots -1}^{N-l},\cdots, \overbrace{1,\cdots,1}^{N}) ,
\end{eqnarray}
which act on the $|n_i\rangle \otimes  |n_{j}\rangle$ space.
If particles of different colors at $i$ and $j$-th sites are exchanged, $Q_{i,j}$ inverts the sign of a state vector.
Thus, it is shown that ${\cal Q}\Gamma^{\rm c}{\cal Q}=-\Gamma^{\rm c}$, ${\cal Q} \Gamma^{\rm r}{\cal Q} = \Gamma^{\rm r}$, and  ${\cal Q}\Gamma^{\rm h}{\cal Q} = \Gamma^{\rm h}$.
We therefore obtain the Hamiltonian,
\begin{equation}
\tilde{\cal H}={\cal Q}\hat{\cal H}{\cal Q} = \sum_i \tilde{h}_{i,i+1}
\label{sonnonegative}
\end{equation}
where
\begin{equation}
\tilde{h}_{i,i+1}\equiv   -\Gamma_{i,i+1}^{\rm c} + \alpha \Gamma_{i,i+1}^{\rm r} - (1-\alpha)\Gamma_{i,i+1}^{\rm h} ,
\end{equation}
which has no negative-sign problem  for $\alpha\le 1$.
If we use the Schwinger boson representation for $\Gamma$s, this Hamiltonian describes $N$-color interacting bosonic particles, as well.

We further examine the correlation function of the SO($N$) chain.
Similarly to the SO(3) case, the result for the conventional spin correlation function is given by
\begin{equation}
\langle L^{ab}_i L_j^{ab} \rangle_{\hat{H}} = - \langle T^{ab}_i e^{i\pi\sum_{i<k<j}T_k^{ab}}T_j^{ab} \rangle_{\tilde{H}} 
\label{soncorrelation}
\end{equation}
where $T^{ab}$ is defined as a symmetric matrix,
\begin{equation}
(T^{ab})_{x,y}=  \delta_{a,x}\delta_{b,y} +\delta_{b,x}\delta_{a,y} .
\end{equation}
Note that we have used the identity $e^{i\pi L^{ab}}=e^{i\pi T^{ab}}$ in Eq. (\ref{soncorrelation}).
It is also straightforward to confirm that the string correlation function is mapped as 
\begin{equation}
\langle L^{ab}_i e^{i\pi \sum_{i<k<j}L_k^{ab}} L_j^{ab} \rangle_{\hat{H}} = - \langle T^{ab}_i T_j^{ab} \rangle_{\tilde{H}}.
\label{eq:so_correlation}
\end{equation}

In the Cartan-generator-diagonal base,  a KT transformation is defined for each Cartan generator\cite{Xiang1, Xiang2}, where  construction of the ${\cal V}$-transformation is highly nontrivial.
In contrast, the generalized Jordan-Wigner transformation maps the string order parameters into the ferromagnetic orders all at once.
In particular, it should be remarked that, although the KT transformation is still unknown  for $N=$even,  the negative sign was removed by the Jordan-Wigner approach independently of $N$.

At the VBS point $\alpha=\frac{N-2}{N}$, the Hamiltonian is the spatial sum of the projection operator to the  SO($N$)-symmetric sector with the dimension $(N+2)(N-1)/2$.\cite{Xiang1, Xiang2}
Thus, the groundstate of Eq. ({\ref{sonh}) can be exactly represented by the MPS, where  the groundstate energy per spin is exactly $-1$.
For the Hamiltonian (\ref{sonnonegative}) where the negative sign is removed, we can easily obtain the groundstate wavefunctions by diagonalizing $\hat{h}_{i,i+1}$:
\begin{equation}
|\Phi^\nu \rangle \equiv \prod_i  |\phi^\nu_i\rangle ,
\label{classical_son}
\end{equation}
where
\begin{equation}
|\phi^\nu_i\rangle =\frac{1}{\sqrt{N}}\sum_{n_i=1}^N \sigma^{\nu}(n_i) |n_i\rangle ,
\label{songsvec}
\end{equation}
with $\sigma^\nu(n_i) = \pm 1$.
Clearly, the state $|\Phi^\nu \rangle$ is the product state with respect to the on-site wavefunction $|\phi^\nu_i\rangle $, where the all spins on the chain are disentangled from each other.
Thus,  we can easily calculate the expectation value of the string order parameter,
\begin{equation}
\langle\phi_i^\nu| T_i^{ab} |\phi_i^\nu\rangle = \pm \frac{2}{N} ,
\end{equation}
for the broken symmetry state of Eq. (\ref{songsvec}).

The number of the degeneracy can be counted as the possible number of $\{\sigma^\nu(n_i) \}$.
Taking account of the irrelevance of the overall sign, we straightforwardly obtain the number of degeneracy as  $2^{N-1}$, so that $\nu$ runs $1,\cdots 2^{N-1}$.
In terms of the MPS for $N=$odd, the groundstate is described by the MPS of a  $2^{(N-1)/2}\times 2^{(N-1)/2}$ matrix, where the number of degeneracy clearly corresponds to that of the edge states attributed to the $(Z_2\times Z_2)^{(N-1)/2}$ symmetry.
For $N=$even, the groundstate is also described by the MPS of a $2^{N/2}\times 2^{N/2}$ matrix, but the half of its matrix elements are zero.
Therefore, it is concluded that the number of the groundstate degeneracy is consistent with the MPS in the Cartan-generator-diagonalizing basis.

Here, we should make a comment on the linear dependence of the groundstate of Eq. (\ref{songsvec}).
For the local Hamiltonian at the VBS point, the groundstate belongs to the SO($N$) singlet sector or the anti-symmetric sector with dimension $N(N-1)/2$.\cite{Xiang1, Xiang2}
Thus there are only $1+N(N-1)/2$ number of linear independent eigenvectors of the groundstate.
This implies that, for $N > 3$,  the vectors $|\phi^\nu_i\rangle $ for $\nu=1\cdots 2^{N-1}$ are linearly dependent within the two-site problem. 
However, when the chain becomes long, the product state $|\Phi^\nu \rangle$ recovers its linear independence, where the total Hilbert space is exponentially enlarged.

\section{quantum Monte Carlo simulation}

In the previous section, we have obtained the negative-sign-free Hamiltonian (\ref{sonnonegative}) for the SO($N$) BLBQ chain, which can be regarded as the problem of the interacting bosonic particles with $N$ colors, on the basis of the Schwinger boson representation.
In this section, we demonstrate QMC simulation  based on a directed loop algorithm for the negative-sign-free Hamiltonian.

\subsection{directed-loop algorithm}
\label{sec:dla}
To formulate a QMC algorithm for the $N$ color particle models of
Eq. (\ref{sonnonegative}), the world-line representation of partition
function is useful(See the review paper Ref. [\onlinecite{Harada::2004}]). Using Suzuki-Trotter decomposition and inserting
the $N$ color Hilbert basis $\vert n \rangle(= \otimes_{i=1}^L \vert n_i \rangle)$ between decomposed
terms, the partition function is written as
\begin{align}
Z&={\rm Tr} \exp(-(\tilde{\cal H}-C)/T) \approx  \sum_{n(\cdot)} W(\{n\}),
\label{eq:world}\\
W(\{n\}) &\equiv \prod_{t=1}^{ML}\prod_{i=1}^L
\langle n(t+1) \vert
(1-(\tilde{h}_{i,i+1}-C)\delta\tau)
\vert n(t) \rangle,
\end{align}
where $T$ is a temperature and $L$ is the chain length and $\vert n(ML+1) \rangle = \vert n(1) \rangle$.
Also,  $\delta\tau = 1/MT$ denotes the imaginary-time slice discritized by the Trotter number $M (\gg 1)$.
We rescale the temperature such that the Boltzmann constant $k_B=1$. 
We set the periodic boundary condition for Trotter direction $t$. 
The classical configuration $n(\cdot)$ can be drawn as the $N$ color world-line configuration in two dimensional space (See Fig. \ref{fig:worldline}). 
The $W(\{n\})$ can be regarded as the classical Boltzmann weight.  
The constant $C$ does not affect the expectation values of physical quantities, where  $C \ge \alpha$ is necessary so that the Boltzmann weight $W(\{n\})$ is positive.
\begin{figure}[tb]
\centering
\includegraphics[width=0.4\textwidth]{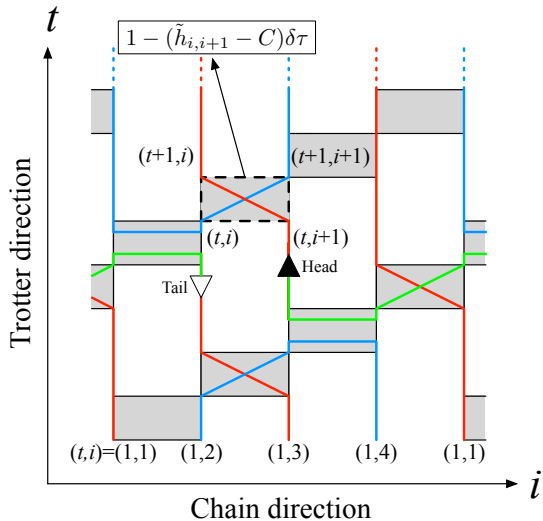}
\caption{$N=3$ color world-line configuration of Eq. \eqref{eq:world} with $L = 4$ with a periodic boundary. The gray box denotes a imaginary time slice with a local Boltzmann weight $\langle
  n(t+1) \vert (1-(\tilde{h}_{i,i+1}-C)\delta\tau) \vert n(t)
  \rangle$. The solid up and down triangles are a worm's head (black) and tail (white). }
\label{fig:worldline}
\end{figure}

We can construct the directed-loop algorithm for this classical model.
It introduces a pair of singular points into the world-line configuration, which are called the worm's head or tail (See the solid up and down triangles in Fig. \ref{fig:worldline}). 
The colors of world lines change at these singular points. 
We first insert a worm's head and tail on a world-line configuration so that the new color is inserted into an initial world-line at random. 
Setting the worm's head and tail  virtually at the same position, we can insert it freely.
We change the world-line configuration by moving these worm's head or tail.
If the worm's head meets its tail, we can annihilate them. 
Then, we obtain a new world-line configuration which may be globally different from the initial configuration. 
In practice, we  move only the worm's head. 
The direction of movement is represented as the up or down triangle in the Trotter direction in Fig. \ref{fig:worldline}.
After the head (black up triangle) hits a slice $\langle n(t+1) \vert (1-(\tilde{h}_{i,i+1}-C)\delta\tau) \vert n(t) \rangle$ from a corner $(t, i+1)$, it will take one of four possible positions: $(t, i), (t, i+1), (t+1, i), (t+1, i+1)$. 
The new direction of the head always points toward the outside of the slice box. The
probabilistic rule for scattering the head satisfies the balance condition between the four possible configurations, because the frequency of a new world-line configuration must be proportional to the classical Boltzmann weight defined without the worm's head and tail. 
The rule which satisfies the balance condition is not unique. 
We use the scatter table determined by Suwa and Todo's reversible kernel of the Markov chain\cite{Suwa::2012}. 
Then, if $C \ge \alpha$, we always obtain the no-turn back table. 
In addition, we take the continuous time limit as $\delta\tau = 0 \ (M \to \infty)$ in the algorithm level. 
Thus, there is no systematic error from the Trotter discretization in our simulations. 
We note that the present algorithm includes the QMC algorithm for the SU($N$) models of Ref. [\onlinecite{Mila}] as a special case.
In addition, the present algorithm can  also be applied to not only one-dimensional model, but also to  higher dimensional models with $N$ color particles. 
It is interesting to mention that a similar QMC simulation was actually performed for a $S=1$ ferromagnetic biquadratic model on a triangular lattice, which includes only the $\Gamma^{\rm h}$ diagram\cite{Kaul}.

\subsection{SO($N$) BLBQ chains at finite temperatures}
\label{sec:finite_temp}
\subsubsection{groundstate phase diagram}

At zero temperature, the SO($N$) BLBQ models of \eqref{eq:SONBLBQ} have four types of phases depending on $N$ and $\alpha$: the Haldane phase for odd $N$, the non-Haldane phase for even $N$\cite{Korezhuk}, the dimer phase, the  ferromagnetic phase, and the critical phase (See Fig. 5 in Ref. [\onlinecite{Xiang1}]). 
In addition, there are a couple of solvable points in the groundstate phase diagram. 
The SU($N$) symmetric point corresponds to $\alpha=1$,  which can be solved by Bethe ansatz and has a gapless groundstate.\cite{Sutherland}
Another Bethe ansatz solvable point is located at $\alpha = (N-4)/(N-2)$, where the groundstate is also gapless\cite{TB,Reshetikhin}. 
The phase between these two Bethe ansatz solvable points is in the Haldane phase or the non-Haldane phase, where the symmetry-protected topological order appears.
In particular, at $\alpha=(N-2)/N$, the groundstate wavefunction is described by a VBS type wavefunction, which is exactly represented by the MPS. 
On the other hand,  the dimer phase appears between $\alpha = (N-4)/(N-2)$ and $-1$.
 Note that these phases are gapful, where the density matrix renormalization group method works  effectively. 
Thus, there are also many studies of the groundstate properties of the SO($N$) BLBQ chains\cite{Korezhuk,Reshetikhin,Xiang1,Xiang2,Alet,Orus,Orus2}. 
However, the finite temperature behavior is not well studied.

\subsubsection{energy}
We derived the $N$-color bosonic particle model of Eq. \eqref{sonnonegative} from the SO($N$) BLBQ model by the generalized Jordan-Wigner transformation.  
This model has no negative-sign problem for $\alpha \le 1$. 
Thus, using the directed-loop algorithm above, we can estimate its temperature-dependent behavior numerically. 
In the following, we will show the results of QMC simulations at finite temperatures.
The typical number of world-line configurations is more than $10^6$.

Figure \ref{fig:energy} shows the energy of the SO(3), SO(4), and SO(5) BLBQ chains at the  VBS and SU($N$) points. 
The system size $L$ is $256$. 
We cannot see the system size dependence in the scale of Fig. \ref{fig:energy}. 
We note that all cases in Fig. \ref{fig:energy} suffer from the negative-sign problem if we directly calculate the original SO($N$) BLBQ models. 
The horizontal dotted lines are the exact values of the groundstate energy. 
The all results of QMC calculations smoothly converge to the exact values of the groundstate energy.  Thus, we numerically confirmed the correctness of our transformation.

Here, we make a comment on the boundary condition in the QMC simulations.
The negative-sign free Hamiltonian is basically derived under the open boundary condition due to the nonlocal transformations.
However, we have in practice performed the simulations in Fig. \ref{fig:energy} under the usual periodic boundary condition, by neglecting the noncocal interaction between the boundary sites of $i=1$ and $L$.
Of course,  the directed loop algorithm also works for the open boundary system and we have actually confirmed that the results for the open boundary are consistent with those for the periodic boundary, which implies that the periodic boundary condition yields no undesirable effect on the QMC simulation.
However, the finite-size dependence of the results for the periodic boundary is much less than those for the open boundary.
Therefore, we basically show the results for the periodic boundary condition below.

\begin{figure}[tb]
\centering
\includegraphics[width=0.5\textwidth]{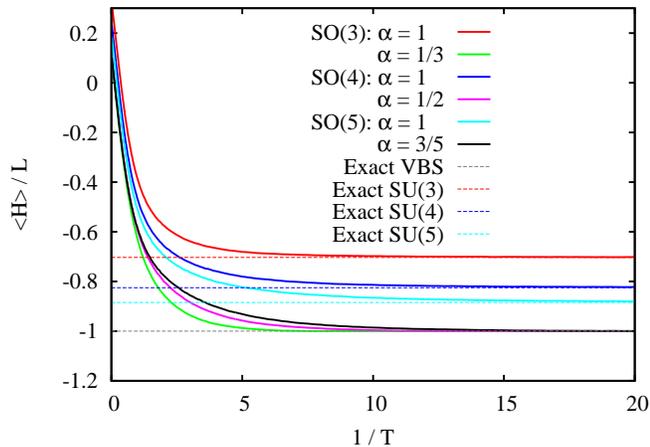}
\caption{Energy per site for SO(3), SO(4), and SO(5) BLBQ models at finite temperatures.
  The system size $L$ is 256. Horizontal dotted lines are the exact values of the groundstate energy.}
\label{fig:energy}
\end{figure}

\subsubsection{entropy}
\begin{figure}[bt]
\centering
\includegraphics[width=0.5\textwidth]{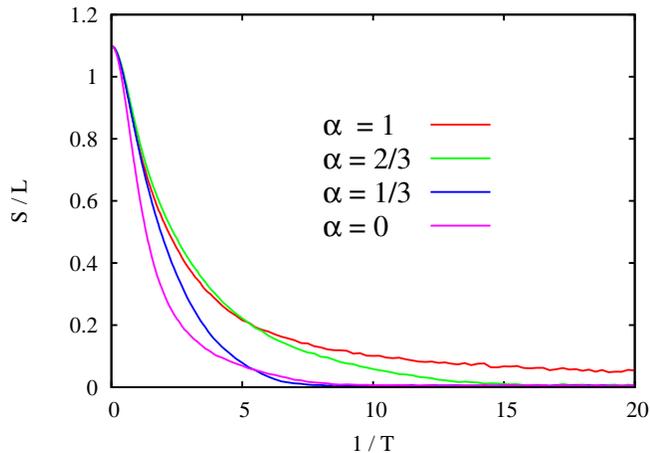}
\caption{Entropy per site of the SO(3) BLBQ model at finite temperatures.
  The system size $L$ is 256.}
\label{fig:entropy}
\end{figure}
If the temperature dependence of energy can be precisely calculated, we can estimate the temperature dependence of thermal entropy $S$ from
a numerical integration of energy as
\begin{equation}
  \label{eq:entropy}
  S(\beta) = S(\beta=0) + \int_0^\beta d\beta' \beta' \frac{\partial \langle \tilde{\cal H} \rangle_{\tilde{\cal H}}}{\partial \beta'},
\end{equation}
where $\beta = 1/T$ and $S(\beta=0)/L = \log(N)$ for the SO($N$) BLBQ models. 
Since the energy rapidly changes in the higher temperature region, we took enough points in the higher temperature regions.
Figure \ref{fig:entropy} shows the entropy of the SO(3) BLBQ models at finite temperatures. In the gapped Haldane phase ($\alpha \ne 1$), the entropy quickly goes to zero when the temperature is lower than the energy gap. 
On the other hand, the entropy of gapless point ($\alpha = 1$) decreases slowly. 
We also found the qualitatively same behavior in the SO(4) and SO(5) models.

As discussed in Ref. [\onlinecite{Mila}], the dependence of entropy growth on the SO($N$) symmetry is important for decreasing the temperature of a system by the evaporative cooling technique. 
If the entropy grows much faster at a low temperature for increasing $N$, the temperature corresponding to a given entropy decreases very fast. 
In Ref. [\onlinecite{Mila}], such behavior was confirmed at the SU($N$) point. 
Here, we have confirmed the faster growth with respect to  $N$ not only at the SU($N$) point, but also in the whole region of $\alpha$. 
Figure \ref{fig:compare_entropy} shows the temperature dependence of entropy at the VBS points of the SO($N$) BLBQ models ($\alpha = (N-2)/N$). 
When $N$ increases, the entropy grows much faster at low temperature. 
At the VBS points, gaps are opened. 
However, even if the temperature is larger than energy gap scale, rapid growth of the entropy is observed. 
The main reason is the size of the Hilbert space for the SO($N$) state.

In the mapped bosonic particle system, the SO($N$) symmetry of the original BLBQ chain is hidden by the nonlocal transformation.
In other words, a bosonic particle model with specific interaction parameters, which does not exhibit  explicit SO($N$) symmetry, can be inversely converted to the SO($N$) BLBQ chain through the nonlocal transformation. 
From the experimental viewpoint for optical lattices, then, an interesting point is that fine tuning of interaction parameters is rather easy compared with usual spin systems.
This suggests that the low-temperature properties of the SO($N$) BLBQ chain may be observed for an optical lattice with efficient evaporative cooling.

\begin{figure}[tb]
\centering
\includegraphics[width=0.5\textwidth]{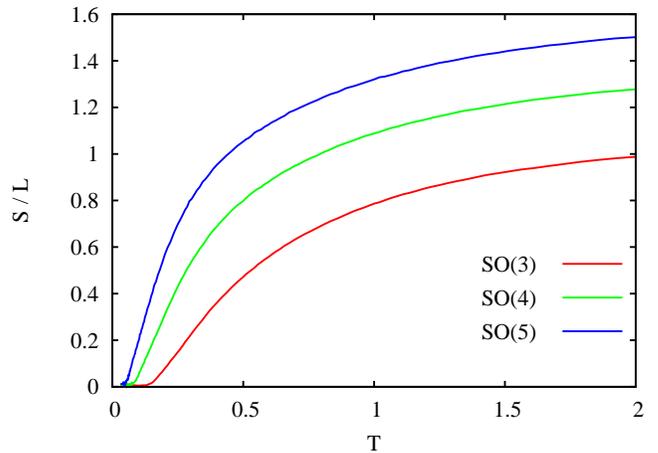}
\caption{Entropy per site of the SO(3), SO(4), and SO(5) BLBQ models at the VBS points ($\alpha = (N-2)/N$).  The system size $L$ is 256.}
\label{fig:compare_entropy}
\end{figure}

\subsubsection{string order parameter}
\label{sec:so}
For  SO($N$) BLBQ chains, the generalized Jordan-Wigner transformation maps the string correlation function into the off-diagonal correlation function $\langle T^{ab}_iT^{ab}_j\rangle_{\tilde{\cal H}}$ defined by Eq. \eqref{eq:so_correlation}.
Then, the hidden-symmetry breaking of the topological order in the original system can be detected as the explicit symmetry breaking characterized by the order parameter $\langle T^{ab}\rangle_{\tilde{\cal H}} $ in the mapped system.


\begin{figure}[tb]
\centering
\includegraphics[width=0.5\textwidth]{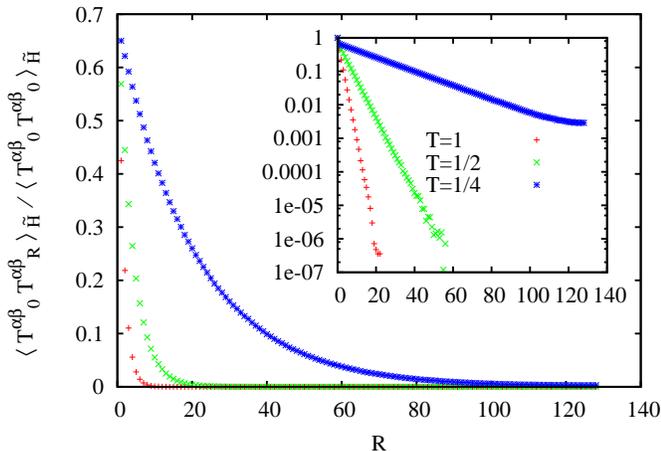}
\caption{Two point correlation function of the off-diagonal order in the mapped system, which corresponds to the string correlation function of the AKLT model. 
The system size $L$ is 256. The inset is a semi log plot.}
\label{fig:correlation}
\end{figure}
\begin{figure}[tb]
\centering
\includegraphics[width=0.5\textwidth]{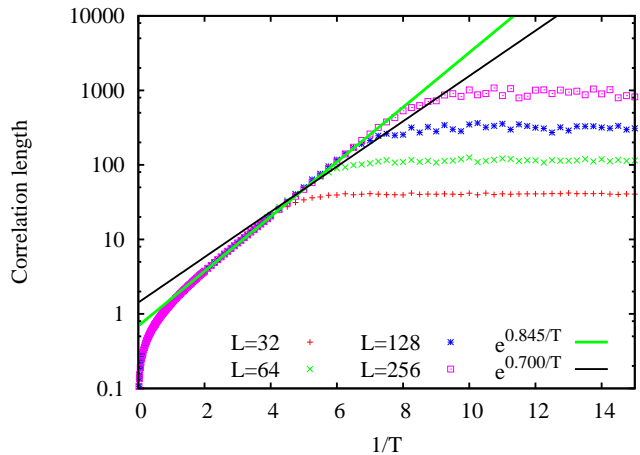}
\caption{Correlation length of the off-diagonal order, which corresponds to the string correlation length at the AKLT point of the $S=1$ BLBQ chain. 
  We use the second-moment of correlation function to estimate the correlation length. }
\label{fig:correlation_length}
\end{figure}

In the QMC simulations, we can measure the two point correlation function of the off-diagonal order at a finite temperature, as follows.
Since the worm's head and tail correspond to the operator $T^{ab}$, the probability such that the worm's head and tail are put in a world-line configuration is proportional to the value of two point Green function, $\langle T^{ab}_i(t)T^{ab}_j(t') \rangle_{\tilde{H}}$. 
Thus, we can estimate the two point correlation function of $T^{ab}$ from the histogram of a distance between head and tail. 
Figure \ref{fig:correlation} shows the two point correlation function at the AKLT point of the SO(3) case, which is estimated from the histogram of the distance between the worm's head and tail. 
The vertical axis of this figure is normalized by $\langle (T^{ab}_0)^2 \rangle_{\tilde{H}}$.
As shown in the inset, the function form can be written as $\langle T^{ab}_iT^{ab}_j\rangle_{\tilde{\cal H}}\propto \exp(-|i-j|/\xi(T))$, where $\xi(T)$ is the correlation length at a temperature $T$. 
Because the AKLT point is in the Haldane phase, the string correlation function grows over the whole distance as the temperature decreases. 
In other words, the correlation length diverges when the temperature goes to zero, which implies that long range order of $T^{ab}$ occurs in the groundstate.
Figure \ref{fig:correlation_length} shows the correlation length estimated from the second-moment of the correlation function, which diverges as $\xi(T) \propto \exp(\Delta_\xi /T)$. 
We estimated $\Delta_\xi$ as $0.845$, using the results between $1/T=4$ and 6. 
Note that, at the AKLT point, the groundstate in the mapped system is reduced to be the classical product state of Eq. (\ref{classical_aklt}).
Thus, we may intuitively expect that $\Delta_\xi$ would coincide with the energy gap from the groundstate, as in the 1D Ising chain. 
However, it is different from the known value $\Delta = 0.700$ which has been evaluated by a variety of methods such as variational approximation\cite{Schraf},  the magnetization curve\cite{OHA}, and the tensor network methods\cite{CL}. 
The value $0.845$ suggests that a nontrivial quantum effect is involved in the excited states, although the groundstate has no entanglement.
We will report further details of the off-diagonal order in the $N$ color particle model elsewhere.


\section{summary and discussions}

In this paper, we have studied the relation between the symmetry-protected topological order and the negative-sign problem for SO($N$) BLBQ chains.
A generalized Jordan-Wigner transformation combined with the defining representation of the SO($N$) spin removes the negative sign in the parameter region corresponding to  VBS-type phases.
So far, the negative-sign problem in QMC simulations has been addressed for such particular models as the zigzag ladder\cite{Nakamura} and SU($N$) chain\cite{Mila}, and there is no systematical approach even in one dimension. 
A key point in this paper is that the generalized Jordan-Wigner transformation is reconstructed from the pair dientanglers, which provides a systematical construction of the nonlocal transformation.
Moreover, the symmetry among the rotational axes  hidden by the Cartan-generator-diagonalzing representation becomes  manifest in the defining representation.
Then, the explicit matrix representation of the transformation enables us to systematically disentangle the symmetry-protected topological entanglement and to obtain the correlation functions in the mapped system.
It is also interesting that this transformation totally disentangles the groundstate entanglement  for $N >3$, in contrast to the fact that a series of  KT transformations is required in order to disentangle the VBS state\cite{Xiang1}.
Recently, the relation between the nonlocal unitary transformation and the symmetry-protected topological order has attracted much attention\cite{Else,Quella}.
Our results might stimulate further investigations of negative-sign problems for a variety of systems associated with symmetry-protected topological orders\cite{SPT}.

We have formulated a directed loop algorithm for the negative-sign-free Hamiltonian, which is nothing but the $N$-color  bosonic particle model.
The scattering of the world lines is given by the diagrams in Fig. \ref{diagram}.
QMC simulations were demonstrated for the SO($N$) BLBQ chains with $N=3, 4, 5$ and their thermodynamic behaviors were clarified.
Although the AKLT model is a most essential model in quantum spin systems, it has been difficult to reveal its thermodynamics, because of the negative sign problem.
In particular, we would like to emphasize that the present approach enables us to obtain the temperature dependence of the string correlation function, which may provides a further insight into the thermodynamics of the hidden topological order.

At the VBS point, the VBS state of the original BLBQ Hamiltonian is particularly mapped into the classical product state of Eq. (\ref{classical_son}).
The number of the degeneracy is consistent with the edge degrees of freedom in the MPS representation in the original system.
The groundstate wavefunction of the mapped system (\ref{classical_son}) is clearly invariant under translation operation, independently of $N$. 
In the spin ladder representation of the SO(4) BLBQ chain, however, the groundstate is interpreted as the staggered dimer pattern accompanying the spontaneous translational symmetry breaking in the leg-direction.\cite{Korezhuk}
A reason for this is that the staggered dimer operator is not SO(4)-invariant and thus becomes nonlocal in the mapped system.
For the connection between the negative-sign problem and the groundstate with  broken translational symmetry, a further investigation of the zigzag ladder or the SO(4) BLBQ chain may provide an interesting perspective.

What we have clarified in this paper is that the symmetry-protected topological entanglement is a possible source of the negative sign, and this negative sign can be removed by the nonlocal transformation.
However, we should finally remark that the sign problem involves information  not only of the groundstate but also of the excited states, which implies that an argument relying only on the  groundstate entanglement may occasionally be misleading.
For instance, the groundstates of the $S=1/2$ zigzag ladder or the SO(4) BLBQ chain at the VBS point are exactly described by the decoupled dimers accompanying the translational symmetry breaking, where there is no entanglement between the decoupled dimers.
Thus, one might expect that a local unitary transformation disentangling the dimer pair could remove the negative sign.
However, this is not the case; the nonlocal transformation was actually required, as was seen in this paper.
In this sense, the hidden symmetry, which is also relevant to the excited states, plays a crucial role behind both of the negative-sign problem and the symmetry-protected topological entanglement.

\begin{acknowledgments}
The authors would like to thank T. Nakamura, H. Katsura, N. Kawashima, and K. Totsuka for valuable comments and discussions.
This work was supported in part by Grants-in-Aid No. 25800221, 23540450, 23540442 and 23340109 from the Ministry of Education, Culture, Sports, Science and Technology of Japan. 
The numerical computations were performed on computers at the Supercomputer Center, ISSP, University of Tokyo, and at the Supercomputer Center of Kyoto University.
\end{acknowledgments}




\begin{thebibliography}{99}


\bibitem{Troyer} M. Troyer and U.-J. Wiese, Phys. Rev. Lett. {\bf 94}, 170201 (2005).

\bibitem{Nakamura} T. Nakamura, Phys. Rev. B, {\bf 57},  R3197 (1998).


\bibitem{Mila}  L. Messio and F. Mila, Phys. Rev. Lett. {\bf 109}, 205306 (2012).

\bibitem{Frischmuth} B. Frischmuth, F. Mila, and M. Troyer, Phys. Rev. Lett. {\bf 82}, 835 (1999).

\bibitem{KT} T. Kennedy and H. Tasaki, Phys. Rev. B {\bf 45}, 304 (1992).

\bibitem{Kennedy} T. Kennedy, J. Phys.:condens. Matter {\bf 6}, 8015 (1994).

\bibitem{Kohmoto-Tasaki} M. Kohmoto and H. Tasaki, Phy. Rev. B {\bf 46} 3486 (1992). 


\bibitem{Sutherland}C. K. Lai, J. Math. Phys. 15, 1675 (1974); B. Sutherland, Phy. Rev. B {\bf 12}, 3795 (1975).


\bibitem{AKLT} I. Affleck, T. Kennedy, E. H. Lieb, and H. Tasaki, Phys. Rev. Lett. {\bf 59}, 799 (1987). 

\bibitem{Wen}Z.-C.Gu and X.-G. Wen  Phys. Rev. B {\bf 80}, 155131 (2009).

\bibitem{Pollmann} F. Pollmann, A. M. Turner, E. Berg, and M. Oshikawa, Phys. Rev. B {\bf 81}, 064439 (2010); 
F. Pollmann, E. Berg, A. M. Turner, and M. Oshikawa, Phys. Rev. B {\bf 85}, 075125 (2012).



\bibitem{Haldane} F. D. M. Haldane, Phys. Lett. A {\bf 93}, 464 (1983); Phys. Rev. Lett. {\bf 50}, 1153 (1983).



\bibitem{Xiang1} H.-H. Tu, G.-M. Zhang and T. Xiang, Phys. Rev. B {\bf 78}, 094404 (2008).
\bibitem{Xiang2} H.-H. Tu, G.-M. Zhang, T. Xiang, Z.-X. Liu, and T.-K. Ng, Phys. Rev. B {\bf 80}, 014401 (2009).


\bibitem{Sandvik} O. F. Syljuasen and Anders W. Sandvik, Phys. Rev. E 66, 046701 (2002).
\bibitem{Harada::2004} N.~Kawashima and K.~Harada, J. Phys. Soc. Jpn. {\bf 73}, 1379 (2004).



\bibitem{Solyom} J. S{\'o}lyom, Phys. Rev. B {\bf 36},  8642 (1987).

\bibitem{Nomura} K. Nomura and S. Takada,  J. Phys. Soc. Jpn. {\bf 60}, 389  (1991).

\bibitem{Fath}G. F{\'a}th and J. S{\'o}lyom, Phys. Rev. B {\bf 44}, 11836 (1991). 
\bibitem{Xiang0}T. Xiang and G. A. Gehring, Phys. Rev. B {\bf 48}, 303 (1993).

\bibitem{Schollweck}U. Schollw{\"oc}k, Th. Jolicoeur, and T. Garel, Phys. Rev. B {\bf 53}, 3304 (1996).

\bibitem{Itoi}C. Itoi and M.-H. Kato, Phys. Rev. B {\bf 55}, 8295 (1997).


\bibitem{Batchelor} M. N. Barber and M. T. Batchelor, Phys. Rev. B {\bf 40}, 4621 (1989).

\bibitem{TB} H. M. Babujan, Nucl. Phys. B215, 317 (1983); L. A. Takhtajan, Phys. Lett. 87A, 479 (1982).

\bibitem{Harada::2001} K.~Harada and N.~Kawashima, J. Phys. Soc. Jpn.{\bf 70}, 13 (2001).

\bibitem{Harada::2002} K.~Harada and N.~Kawashima, Phys. Rev. B {\bf 65}, 052403 (2002).


\bibitem{Nijs} M. den Nijs and K. Rommelse, Phys. Rev. B {\bf 40}, 4709 (1989).


\bibitem{Oshikawa} M. Oshikawa, J. Phys.:condens.matt. {\bf 4}, 7469 (1992).

\bibitem{Okunishi} K. Okunishi, Phys. Rev. B {\bf 83}, 104411 (2011).


\bibitem{TLalgebra} If $\alpha\to \infty$, the world-line diagrams reduce to only $\Gamma^{\rm c}$, which corresponds to that of Temperly-Lieb algebra.



\bibitem{MPSAKLT} M. Fannes, B. Nachtergaele, and R. W. Werner, Commun. Math. Phys. {\bf 144}, 443 (1992). 
A.Kl\"umper, A.  Schadschneider, and J. Zitterz, Eur. Phys. Lett. {\bf 24}, 293 (1993).

\bibitem{Georgi} H. Georgi, {\it Lie Algebras in Particle Physics} (Perseus Biiks, Reading, MA. 1999).




\bibitem{Suwa::2012} H.~Suwa and S.~Todo, Monte Carlo Methods and Applications: Proceedings of the 8th IMACS Seminar on Monte Carlo Methods, 213 (2012); H.~Suwa and S.~Todo, arXiv:1106.3562.


\bibitem{Kaul} R. K. Kaul, Phys. Rev. B {\bf 86}, 104411 (2012).



\bibitem{Korezhuk} A. K. Kolezhuk and H.-J. Mikeska, Phys. Rev. Lett. {\bf 80}, 2709 (1998).




\bibitem{Reshetikhin}
N. Y. Reshetikhin, Lett. Math. Phys. {\bf 7}, 205 (1983) Theor. Math. Phys. {\bf 63}, 555 (1985).



\bibitem{Alet} F. Alet, S. Capponi, H. Nonne, P. Lecheminant, and I. P. McCulloch, Phys. Rev. B {\bf 83}, 060407(R) (2011).

\bibitem{Orus}H.-H. Tu and R. Or{\`u}s, Phys. Rev. Lett. {\bf 107}, 077204 (2011).

\bibitem{Orus2} R. Or{\'u}s, T.-C. Wei, and H.-H. Tu, Phys. Rev. B {\bf 84}, 064409 (2011).



\bibitem{Schraf}R. Scharf and H.-J. Mikeska, J. Phys.: Condens. Matter {\bf 7}, 5083 (1995).

\bibitem{OHA} K. Okunishi, Y. Hieida and Y. Akutsu. Phys. Rev. B {\bf 59}, 6806 (1999).

\bibitem{CL} A.~Garcia-Saez, V.~Murg, T.C.~Wei, arXiv:1308.3631 [cond-mat.str-el].



\bibitem{Else} D. V. Else, S. D. Bartlett, and A. C. Doherty, Phys. Rev. B {\bf 88}, 085114 (2013).

\bibitem{Quella} K. Duivenvoorden and T. Quella, Phys. Rev. B {\bf 88}, 125115 (2013).


\bibitem{SPT}Z. Nussinov and G. Ortiz, Ann. Phys. {\bf 324}, 977 (2009).

\end{thebibliography}
\end{document}